\documentclass[acmsmall,screen,review]{acmart}

\setcopyright{none}
\renewcommand\footnotetextcopyrightpermission[1]{}

\AtBeginDocument{%
  }

\usepackage{amsmath,amsfonts}

\usepackage{algorithm}
\usepackage{algpseudocode}

\usepackage{needspace}
\usepackage{array}
\usepackage{multirow}
\usepackage{tabularx}
\usepackage{threeparttable}
\usepackage{diagbox}
\usepackage{subcaption}
\usepackage[table]{xcolor}
\usepackage{listings}
\usepackage{placeins}
\usepackage{makecell}
\usepackage{xspace}
\usepackage{enumitem}
\usepackage{multirow}

\newcolumntype{L}[1]{>{\raggedright\arraybackslash}m{#1}}
\newcolumntype{C}[1]{>{\centering\arraybackslash}m{#1}}
\newcolumntype{Y}{>{\raggedright\arraybackslash}X}
\providecommand{\NoCaseChange}[1]{#1}

\newcommand{\tool}{\NoCaseChange{ReproShift}\xspace}

\definecolor{codebg}{RGB}{248,248,248}
\definecolor{codetext}{RGB}{70,70,70}
\definecolor{codekeyword}{RGB}{180,30,150}
\definecolor{codecomment}{RGB}{60,145,85}
\definecolor{codestring}{RGB}{50,100,120}
\definecolor{codenumber}{RGB}{120,120,120}
\definecolor{coderule}{RGB}{190,190,190}

\definecolor{deepred}{RGB}{153,0,0}
\definecolor{deepblue}{RGB}{0,51,153}
\definecolor{deepgreen}{RGB}{0,110,60}
\begin{document}
\raggedbottom
\title{\tool{}: Cross-Version Adaptation of Linux Kernel Fuzzing Crash Reproducers}

\lstdefinelanguage{syz}{
  morekeywords={
    syz_mount_image,
    open,
    openat,
    pwritev2,
    ftruncate,
    mmap,
    sendfile,
    ioctl,
    writev
  },
  sensitive=true,
  morestring=[b]",
  morecomment=[l]{//},
}

\lstdefinestyle{syzstyle}{
  language=syz,
  basicstyle=\ttfamily\footnotesize\color{codetext},
  keywordstyle=\color{codekeyword},
  commentstyle=\color{codecomment},
  stringstyle=\color{codestring},
  numbers=left,
  numberstyle=\scriptsize\color{codenumber},
  numbersep=7pt,
  stepnumber=1,
  backgroundcolor=\color{codebg},
  frame=leftline,
  rulecolor=\color{coderule},
  framesep=5pt,
  framexleftmargin=8pt,
  xleftmargin=16pt,
  showstringspaces=false,
  showspaces=false,
  showtabs=false,
  keepspaces=true,
  columns=fullflexible,
  breaklines=true,
  breakatwhitespace=false,
  tabsize=2,
  aboveskip=4pt,
  belowskip=4pt
}

\author{Changjian Liu}
\email{LiuChangjian@csu.edu.cn}
\orcid{0009-0009-8132-4384}
\affiliation{%
  \institution{Central South University}
  \city{Changsha}
  \country{China}
}

\author{Zhiwen Tang}
\email{tang\_zw@csu.edu.cn}
\orcid{0009-0002-7954-3800}
\affiliation{%
  \institution{Central South University}
  \city{Changsha}
  \country{China}
}

\author{Xianying Wang}
\email{wangxianying@csu.edu.cn}
\orcid{0009-0000-5992-1611}
\affiliation{%
  \institution{Central South University}
  \city{Changsha}
  \country{China}
}

\author{Fan Ding}
\email{dingdang66686@csu.edu.cn}
\orcid{0009-0007-3889-1004}
\affiliation{%
  \institution{Central South University}
  \city{Changsha}
  \country{China}
}

\author{Zhengshu Wang}
\email{wangzhengshu39@gmail.com}
\orcid{0009-0009-3879-2844}
\affiliation{%
  \institution{Central South University}
  \city{Changsha}
  \country{China}
}

\author{Chao Hu}
\email{huchao@csu.edu.cn}
\affiliation{%
  \institution{Central South University}
  \city{Changsha}
  \country{China}
}

\author{Yuheng Shen}
\email{syh1308@gmail.com}
\orcid{0000-0002-2667-5431}
\affiliation{%
  \institution{Tsinghua University}
  \city{Beijing}
  \country{China}
}

\author{Ying Fu}
\email{fuying1995@foxmail.com}
\orcid{0000-0001-5329-2586}
\affiliation{%
  \institution{Tsinghua University}
  \city{Beijing}
  \country{China}
}

\author{Heyuan Shi}
\authornote{Corresponding author.}
\email{hey.shi@foxmail.com}
\orcid{0000-0002-9040-7247}
\affiliation{%
  \institution{Central South University}
  \city{Changsha}
  \country{China}
}


\begin{abstract}
Reliable crash reproduction is essential for diagnosing and repairing
kernel crashes. Historical crash reproducers preserve concrete bug-triggering
knowledge and can be reused for regression testing, bug revalidation,
and impact analysis across kernel versions. However, kernel evolution
can invalidate these reproducers by changing kernel interfaces,
configurations, and execution paths. Migrating them is therefore challenging: the adapted reproducer must identify and modify version-sensitive behavior while preserving the bug-triggering operations and resource dependencies that define the original crash.
To address this challenge, we propose \tool{}, an automated framework for cross-version crash reproducer migration by preserving bug-triggering semantics while adapting only version-sensitive behavior. Given source crash artifacts, a source crash reproducer, and a target kernel, \tool{} produces a migrated crash reproducer with same-bug validation evidence.
Specifically, \tool{} first constructs a structured source-to-target
context to identify migration-relevant kernel changes. It then performs
constrained target-aware adaptation to revise version-sensitive
reproducer operations while preserving the original trigger logic.
Finally, it executes the adapted candidates on the target kernel and uses
execution feedback and case-specific crash evidence to iteratively
validate candidates and refine subsequent migrations.
In a real-world deployment on an evolving upstream kernel, \tool{}
further produces 22 upstream crash reports generated from adapted
candidates, of which 7 have been fixed and 5 have received explicit
maintainer confirmation. To systematically quantify its migration effectiveness,
we construct a benchmark of 107 real-world source--target migration cases. \tool{} achieves confirmed same-bug migration in 57 cases, compared with 19 cases using direct replay.
\end{abstract}





\maketitle
\pagestyle{plain}
\thispagestyle{plain}

\section{Introduction}

Modern Linux kernels evolve continuously across upstream mainline,
stable, long-term-support, and vendor-specific branches, making kernel
testing and maintenance increasingly challenging. Kernel fuzzing has
become an important technique for exposing defects that are difficult
to discover through conventional testing. Large-scale fuzzing
infrastructures continuously produce crash reports, crash reproducers, kernel
configurations, and execution artifacts
~\cite{Syzkaller,Syzlang,syzbot}. These artifacts preserve concrete
bug-triggering knowledge and provide an important basis for subsequent
debugging, patch validation, regression testing, and impact analysis.

The value of these artifacts depends on whether their bug-triggering
knowledge remains reusable as the kernel evolves. Such reuse can proceed
in either chronological direction. A crash reproducer obtained from an older
kernel may need to be reused on a newer branch for bug revalidation,
regression testing, or debugging after substantial kernel evolution.
Conversely, a crash reproducer generated on a newer upstream kernel may need
to be applied to older stable, long-term-support, or vendor kernels to
determine whether the same defect affects actively maintained branches.
In both scenarios, an existing crash reproducer provides concrete knowledge
of how a defect was previously reached and can potentially avoid
rediscovering the same behavior from scratch. However, kernel evolution
may invalidate the version-specific assumptions encoded in the
reproducer, preventing this knowledge from being directly transferred
to another environment.

Existing kernel testing techniques provide limited support for such
cross-version reuse. Prior work improves seed selection, syscall
modeling, and bug-triggering effectiveness
~\cite{pailoor2018moonshine,sun2021healer,sun2022ksg,SyzVegas}, while directed,
interface-aware, and concurrency-oriented fuzzers explore specific
kernel components and execution paths
~\cite{Hawkeye,tan2023syzdirect,corina2017difuze,jeong2019razzer,
xu2019janus,xu2020krace}. Recent LLM-based techniques further assist
input and specification generation
~\cite{xia2024fuzz4all,yang2025kernelgpt}. Post-discovery approaches
analyze the impact of exposed crashes or reproduce known kernel bugs
~\cite{zou2022syzscope,li2025yome}, but do not address how an existing crash reproducer
should be adapted across evolving kernel versions while preserving
same-bug identity.
In practice, an existing crash reproducer is commonly reused through direct replay. Direct replay assumes that the conditions
required by the original trigger remain valid in the target
environment, but this assumption can break as the kernel evolves. The
reproducer may become invalid, terminate without reaching the original
fault, or follow a different execution path and trigger an unrelated
crash. Moreover, replay failure does not necessarily indicate that the
underlying bug is absent from the target kernel; the known trigger may
simply be incompatible with the target environment. Repeated replay or
rediscovery through general fuzzing therefore provides neither an
efficient nor a reliable mechanism for reusing previously accumulated
crash knowledge.

\begin{figure*}[t]
    \centering

    \begin{subfigure}[t]{0.325\textwidth}
        \centering
        \includegraphics[
            width=\linewidth,
            height=3.0cm
        ]{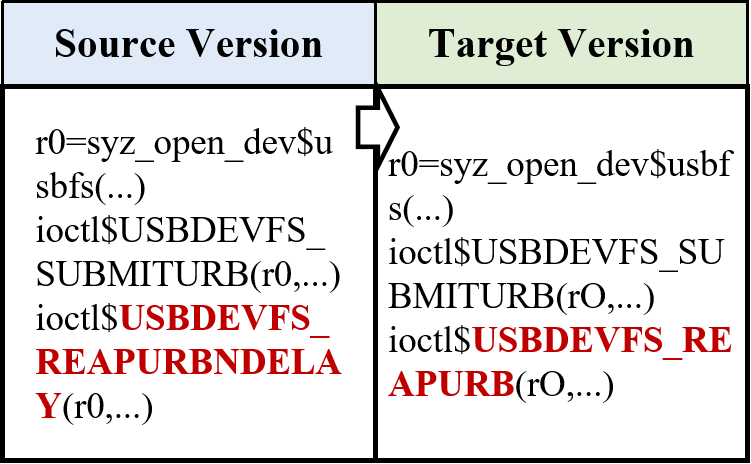}
        \caption{Interface evolution}
        \label{fig:interface-evolution}
    \end{subfigure}
    \hfill
    \begin{subfigure}[t]{0.305\textwidth}
        \centering
        \includegraphics[
            width=\linewidth,
            height=3.0cm
        ]{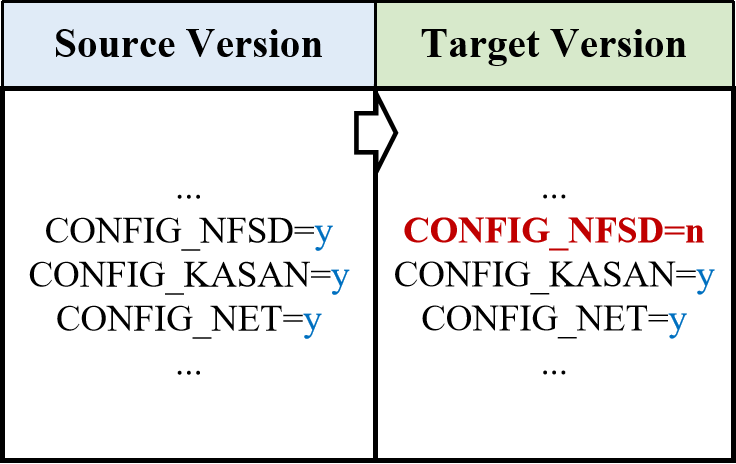}
        \caption{Configuration drift}
        \label{fig:configuration-drift}
    \end{subfigure}
    \hfill
    \begin{subfigure}[t]{0.345\textwidth}
        \centering
        \includegraphics[
            width=\linewidth,
            height=3.0cm
        ]{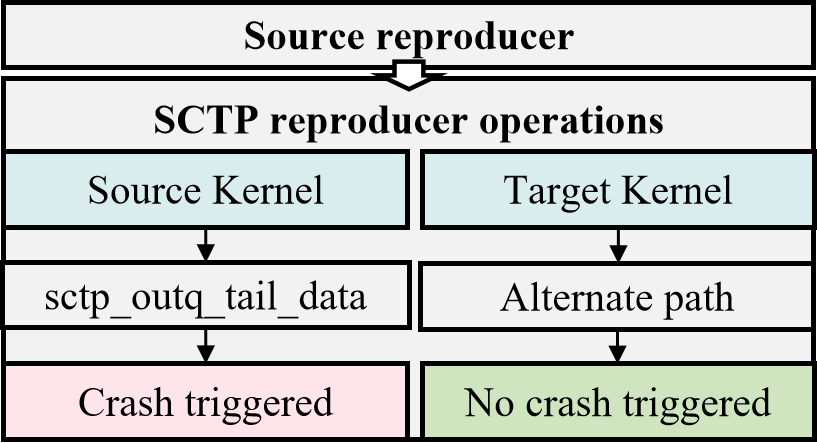}
        \caption{Resulting path divergence}
        \label{fig:execution-path-divergence}
    \end{subfigure}

    \caption{Characteristics of cross-version reproducer drift.
    Interface evolution and configuration drift may invalidate the
    assumptions encoded in a historical crash reproducer, causing its
    execution to diverge from the original bug-triggering path.}
    \label{fig:reproducer-drift}
\end{figure*}

We formulate this problem as \emph{cross-version crash reproducer migration}:
adapting an existing crash reproducer from a source kernel revision to a
different target revision while preserving the identity of the original
kernel crash. This formulation is direction-agnostic: the target
revision may be either later or earlier than the source revision.
Figure~\ref{fig:reproducer-drift} summarizes the main characteristics
observed in cross-version migration cases. As shown in
Figures~\ref{fig:interface-evolution} and
\ref{fig:configuration-drift}, kernel evolution may change the
interfaces and configurations on which a crash reproducer depends. These changes can make operations in the crash reproducer unavailable or alter their runtime behavior, causing execution to diverge from the original
bug-triggering path, as illustrated in
Figure~\ref{fig:execution-path-divergence}. Consequently, a historical
crash reproducer may either fail to execute on the target kernel or execute
without reaching the original kernel crash. These observations highlight two key challenges for reliable cross-version crash reproducer migration.

\textbf{Challenge 1: Kernel evolution makes existing crash reproducers
difficult to migrate.}
A crash reproducer encodes both the operations essential to triggering
the original bug and version-specific assumptions about kernel
interfaces, configurations, resources, and execution behavior.
Directly replaying it on another revision may therefore result in an
invalid program, unavailable functionality, or execution-path
divergence. However, indiscriminately modifying the reproducer may
destroy the original trigger logic. Successful migration must
distinguish version-sensitive behavior that requires adaptation from
bug-triggering operations and resource dependencies that must be
preserved, and then restore a feasible execution path to reproduce the original kernel crash on the target revision.

\textbf{Challenge 2: Cross-version crash reproducer outcomes are difficult
to evaluate.}
Successful execution does not by itself establish successful migration.
An adapted candidate may execute without triggering a crash, or it may trigger an unrelated target-side failure. Moreover, kernel evolution
may change the manifestation, location, stack structure, and execution
context of the same underlying bug. Migration evaluation therefore
cannot rely solely on execution validity, the presence of a crash, or
exact crash-signature matching. It requires case-specific evidence that
connects the observed target-side failure to the original kernel crash,
including compatible failure characteristics, cross-version
crash-anchor correspondence, and bug-relevant execution context.

To address these challenges, we present \tool{}, a target-aware
framework for cross-version crash reproducer migration and validation. First,
\tool{} constructs a structured source-to-target migration context to
identify bug-preserving behavior and migration-relevant kernel changes.
Second, it employs a constrained domain-specific agent to adapt
version-sensitive operations while preserving the original trigger
logic and resource dependencies. Finally, it executes each candidate in
the target environment and validates the resulting evidence using a
case-specific same-bug oracle. Unsuccessful outcomes are classified and
fed back to guide subsequent refinement. This closed-loop design
transforms reproducer reuse from one-shot replay into an iterative
migration-and-validation process.
To evaluate \tool{}, we construct a cross-version crash reproducer migration
benchmark comprising 107 directly comparable source--target cases
derived from real-world Linux kernel crashes. On this benchmark,
\tool{} achieves 57 confirmed same-bug migrations, compared with 19
for direct replay, 18 for direct LLM adaptation, and 17 for direct LLM
adaptation with static self-repair. In a deployment on Linux~7.2-rc3, \tool{} further produces 22 upstream
crash reports generated from adapted candidates, of which 7 have been
fixed and 5 have received explicit maintainer confirmation.

In summary, this paper makes the following contributions:

\begin{itemize}[leftmargin=15pt]

\item We introduce cross-version crash reproducer migration as an important
but underexplored problem in kernel testing and maintenance. We
formulate it as a direction-agnostic task that adapts an existing crash reproducer between kernel revisions, in either the forward or backward direction, while preserving the identity of the original bug.

\item We design and implement \tool{}, a target-aware reproducer
migration framework that combines structured cross-version context
construction, constrained candidate adaptation, case-specific same-bug
validation, and execution-guided refinement to transfer bug-triggering
behavior across kernel environments.

\item We construct a cross-version reproducer migration benchmark
comprising 107 real-world source--target cases and use it to
systematically evaluate \tool{}. The framework achieves a confirmed
same-bug migration rate of 53.3\% (57/107), compared with 17.8\% for
direct replay, 16.8\% for direct LLM adaptation, and 15.9\% for direct
LLM adaptation with static self-repair. A deployment on Linux 7.2-rc3 additionally produces 22 upstream reports generated from adapted candidates, including 7 fixed and 5 maintainer-confirmed cases.

\end{itemize}

\section{Background}
\label{sec:background}
\textbf{Fuzzing Crash Reproduction.}
Kernel fuzzing continuously generates and executes test inputs to expose
faults in complex kernel interfaces and execution paths. When an
execution triggers a warning, sanitizer report, panic, hang, or other
abnormal behavior, modern fuzzing infrastructures preserve the
corresponding crash artifacts for subsequent analysis~\cite{Syzkaller,
Syzlang,syzbot}. These artifacts typically include crash reports,
runtime logs, kernel versions, configurations, and crash reproducers
that record the operations associated with the failure. Together, they
provide both the observed failure evidence and the execution conditions
under which the crash was originally triggered.
A crash reproducer provides an executable representation of a previously
observed failure. It captures the operations, resource dependencies, and
execution assumptions required to reach the original bug. Replaying
these artifacts allows developers to repeatedly reproduce known failures
without rediscovering them through fuzzing from scratch. Crash
reproducers therefore serve as important testing assets for crash
diagnosis, patch validation, regression testing, and other downstream
kernel maintenance tasks.

\textbf{Kernel Evolution and Reproducer Compatibility.}
Kernel evolution continuously changes the execution environment assumed
by historical crash reproducers. Linux evolves simultaneously through
upstream development, stable releases, long-term maintenance branches,
and vendor-specific modifications. Across kernel revisions, interfaces,
configurations, subsystem implementations, and runtime behaviors may
change independently. For example, syscall interfaces and arguments may
evolve, configuration options may enable or disable relevant code paths,
and resource dependencies or interface semantics may differ between
kernel revisions.
These changes directly affect whether a crash reproducer obtained from
one source revision remains compatible with a different target revision.
A crash reproducer encodes operations, resource dependencies, and
execution assumptions whose validity depends on the corresponding kernel
environment. Consequently, compatibility has two dimensions: the
reproducer must first remain executable in the target environment, and
its execution must still preserve the conditions required to reach the
original failure. A crash reproducer can therefore execute to completion
while following a different kernel path, just as it can become invalid
before reaching the intended behavior.

\textbf{Crash Identity and Reproduction Validation.}
Crash reproduction requires determining not only whether a kernel failure occurs, but also which failure has been reproduced. A single test program may exercise multiple kernel components and can therefore expose different failures under different kernel revisions or configurations. The presence of a warning, panic, or sanitizer report alone is insufficient to establish that two executions correspond to the same underlying defect. Kernel crash reports provide additional evidence for distinguishing failures, including crash type, faulting function, stack trace, source location, and other failure-specific information.
Crash identity can therefore be viewed as a relation between the original failure and the failure observed during a later reproduction attempt. In the simplest case, matching crash characteristics provide evidence that the same faulty behavior has been reached. In more complex cases, stack traces or source locations may change across versions even when the underlying defect remains related, while similar crash types may originate from entirely different root causes. Reproduction validation must therefore consider case-specific crash evidence rather than relying solely on a binary crash/no-crash outcome. This distinction is particularly important when historical crash reproducers are reused across evolving kernel environments.

\begin{figure*}[t]
\centering
\includegraphics[width=\textwidth]{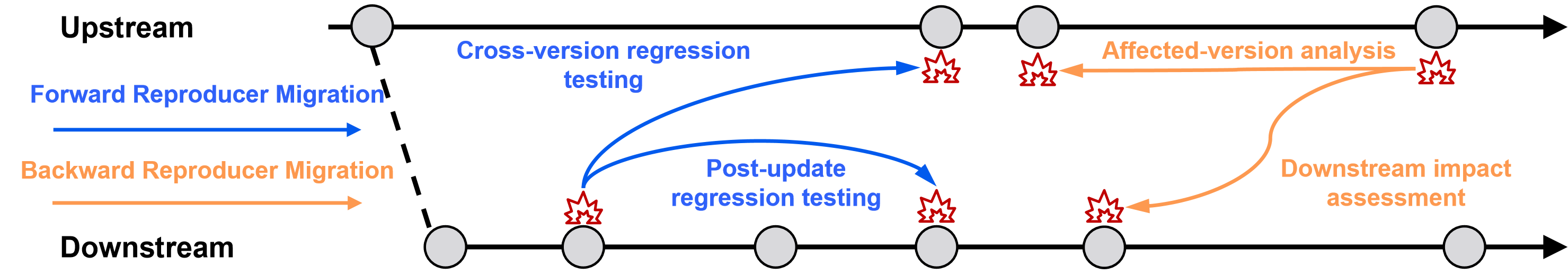}
\caption{Representative maintenance scenarios motivating forward and
backward reproducer migration across evolving kernel branches.}
\label{fig:reuse-motivation}
\end{figure*}

\textbf{Cross-Version Reproducer Reuse.}
\label{sec:reuse-motivation}
Kernel crash reproducers are valuable testing assets whose usefulness
should extend beyond the revision on which they were obtained.
Depending on the maintenance scenario, a crash reproducer may be transferred between earlier and later revisions within the same development lineage or across different maintenance branches. We use \emph{revision} to denote
a concrete kernel commit and \emph{kernel family} to denote its
major--minor release line, such as Linux~5.15, 6.1, or 6.6. We use
\emph{cross-version migration} as the general term for transferring a
crash reproducer between revisions, regardless of migration direction
or family relation. A \emph{branch} denotes a development or
maintenance lineage, such as an upstream or stable branch.
Figure~\ref{fig:reuse-motivation} uses branches to illustrate
representative maintenance scenarios, whereas our evaluation groups
migration pairs by kernel-family relation.

\begin{itemize}[leftmargin=1.5em, itemsep=0.5em, topsep=0.3em]

\item \textbf{Forward migration.}
Forward migration transfers a crash reproducer from an earlier source
revision to a later target revision. The source and target may belong
to either the same or different kernel families. As illustrated by the
blue arrows in Figure~\ref{fig:reuse-motivation}, forward migration
supports two representative maintenance scenarios:
\emph{cross-version regression testing}, which reuses a historical
reproducer on a newer upstream revision, and \emph{post-update
regression testing}, which revalidates known failures after a
downstream branch is updated. More generally, forward migration
preserves previously accumulated bug-triggering knowledge for
regression testing, bug revalidation, and debugging during continuous
kernel evolution. Kernel changes across revisions may modify
interfaces, configurations, resource dependencies, or execution paths,
causing direct replay to fail. Forward migration therefore adapts
version-sensitive behavior introduced during kernel evolution while
preserving the original bug-triggering logic.

\item \textbf{Backward migration.}
Backward migration transfers a crash reproducer from a later source
revision to an earlier target revision. The two revisions may belong
to the same or different kernel families. As illustrated by the orange
arrows in Figure~\ref{fig:reuse-motivation}, this direction can provide
\emph{positive evidence for affected-version analysis}: reproducing
the same bug on an earlier revision establishes that the defect is
reachable there under the tested configuration. It can also inform
downstream impact assessment and backport prioritization. Conversely,
failure to migrate or replay a reproducer does not establish that the
earlier revision is unaffected, because the trigger may rely on
interfaces or execution conditions unavailable on that revision.

\end{itemize}
Although forward and backward migration serve different maintenance
goals, they share the same underlying technical abstraction: adapting a known bug-triggering program from a source revision to a different target revision while preserving the identity of the original bug. The migration direction changes the chronological relationship between the source and target revisions, but not the fundamental requirements of target-aware adaptation and same-bug validation.

\section{Approach}
\label{sec:approach}

\begin{figure*}[t]
\centering
\includegraphics[width=\textwidth]{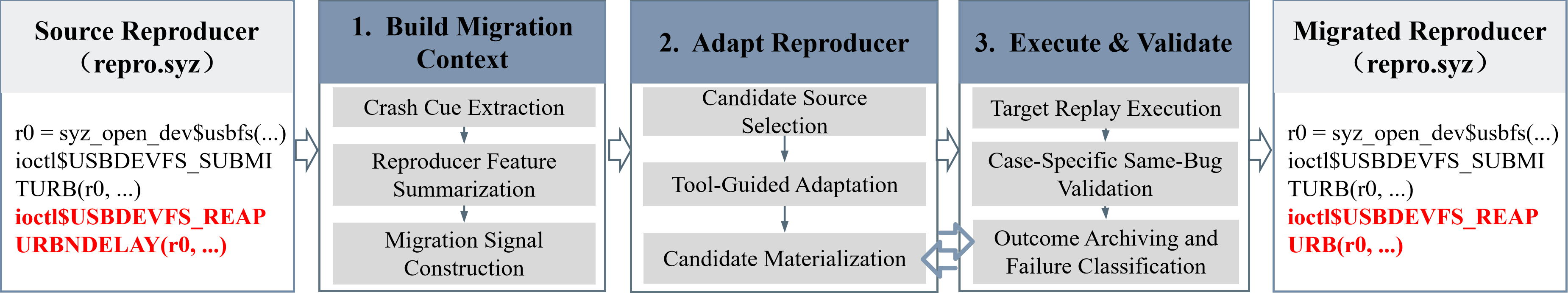}
\caption{Overall workflow of \tool{}.}
\label{fig:architecture}
\end{figure*}

Figure~\ref{fig:architecture} presents the overall workflow of \tool{}.
Given source crash artifacts, a source crash reproducer, and a target kernel,
\tool{} operates in three stages. First, it extracts the source crash
semantics, identifies relevant source--target version differences, and
builds a structured migration context. Second, it selects the available
reproducer format, modifies version-sensitive operations while
preserving the original trigger, and generates a candidate reproducer.
Third, it executes the candidate on the target kernel, validates
same-bug identity, and records the resulting outcome. Unsuccessful
outcomes are returned as feedback for subsequent adaptation. Once
same-bug identity is confirmed, \tool{} outputs the migrated
reproducer.

\subsection{Build Migration Context}

\tool{} begins by collecting available source and target artifacts,
including crash reports, runtime logs, reproducer files, kernel
configurations, and metadata. These heterogeneous artifacts are organized
into a unified source--target case representation for subsequent
migration. Missing artifacts are explicitly recorded rather than
discarded, allowing \tool{} to preserve partial evidence during context
construction. Source and target reproducers are maintained separately
even when both are available locally. Target reference reproducers,
target reference crashes, and fixing patches are retained only for
offline evaluation; they are excluded from migration-context
construction and never enter the online feedback loop.

\begin{table*}[t]
\centering
\caption{Evidence used to construct the structured migration context.
Bug-preserving evidence captures the original failure and trigger,
whereas target-sensitive evidence captures kernel and environment
differences requiring adaptation.}
\label{tab:migration-context-evidence}

\scriptsize
\setlength{\tabcolsep}{4pt}
\renewcommand{\arraystretch}{1.15}

\begin{tabularx}{0.92\textwidth}{
    @{}
    >{\centering\arraybackslash\hsize=0.90\hsize}X
    >{\centering\arraybackslash\hsize=1.10\hsize}X
    >{\centering\arraybackslash\hsize=0.90\hsize}X
    >{\centering\arraybackslash\hsize=1.10\hsize}X
    @{}
}
\toprule
\multicolumn{2}{c}{\textbf{Bug-Preserving Evidence}} &
\multicolumn{2}{c}{\textbf{Target-Sensitive Evidence}} \\
\cmidrule(lr){1-2}
\cmidrule(lr){3-4}

\textbf{Crash Report \& Log} &
\textbf{Source crash Reproducer} &
\textbf{Kernel Source \& Configs} &
\textbf{Target Environment} \\
\midrule

\begin{tabular}[t]{@{}c@{}}
\path{kernel_BUG_at}\\
\path{lib/list_debug.c:58}\\
\ldots
\end{tabular}
&
\begin{tabular}[t]{@{}c@{}}
\path{syz_open_dev$usbfs()}\\
\path{USBDEVFS_SUBMITURB}\\
\path{USBDEVFS_REAPURBNDELAY}
\end{tabular}
&
\begin{tabular}[t]{@{}c@{}}
\path{CONFIG_NFSD=y}\\
\path{CONFIG_KASAN=y}\\
\path{CONFIG_NET=y}
\end{tabular}
&
\begin{tabular}[t]{@{}c@{}}
\path{REAPURB} \textcolor{deepgreen}{available}\\
\path{REAPURBNDELAY} \textcolor{deepred}{unavailable}\\
\ldots
\end{tabular}
\\

\bottomrule
\end{tabularx}
\end{table*}
Given the collected artifacts, \tool{} constructs a structured migration
context that separates bug-preserving evidence from target-sensitive
evidence. Table~\ref{tab:migration-context-evidence} summarizes the four evidence sources used to construct this context. Instead of directly concatenating heterogeneous
artifacts as raw text, \tool{} extracts structured signals that describe
which behaviors should be preserved, which behaviors may require
target-side adaptation, and which evidence is required for online
validation.
Bug-preserving evidence is extracted from source-side artifacts and
captures the information required to maintain the identity of the
original defect. Crash reports and runtime logs contribute
\emph{crash cues}, including normalized bug type, faulting function or
crash anchor, stack cues, and other case-specific identifiers used for
same-bug validation. The source crash reproducer contributes
\emph{reproducer features}, including setup operations, resource flow,
trigger operations, argument dependencies, and execution assumptions
associated with the original failure.

Target-sensitive evidence describes the environmental differences that
may affect crash-reproducer migration. Kernel source code,
configurations, and target metadata are analyzed to extract
\emph{migration signals}, including interface differences, path
divergence, configuration-dependent behavior, resource-semantic
changes, feature availability, and target-side reachability cues. Based
on these signals, \tool{} derives an \emph{online oracle configuration}
from source-side crash evidence and target-environment information.
This configuration specifies how candidate executions should be
interpreted during migration without consulting hidden target
references.

We represent the resulting structured context as
$X=\langle C_{\mathrm{bug}}, F_{\mathrm{repro}},
S_{\mathrm{migration}}, \Omega_{\mathrm{online}}\rangle$, where
$C_{\mathrm{bug}}$ denotes the extracted crash cues,
$F_{\mathrm{repro}}$ denotes the crash-reproducer features,
$S_{\mathrm{migration}}$ denotes the source--target migration signals,
and $\Omega_{\mathrm{online}}$ denotes the online oracle configuration.
The resulting context specifies which behavior should be preserved,
which behavior requires target-side adaptation, and which evidence is
needed for an online same-bug decision. Neither the target-reference
crash reproducer, the target-reference crash, nor the fixing patch is
included in $X$.

\subsection{Adapt Reproducer}

\begin{figure*}[t]
\centering
\includegraphics[width=\textwidth]{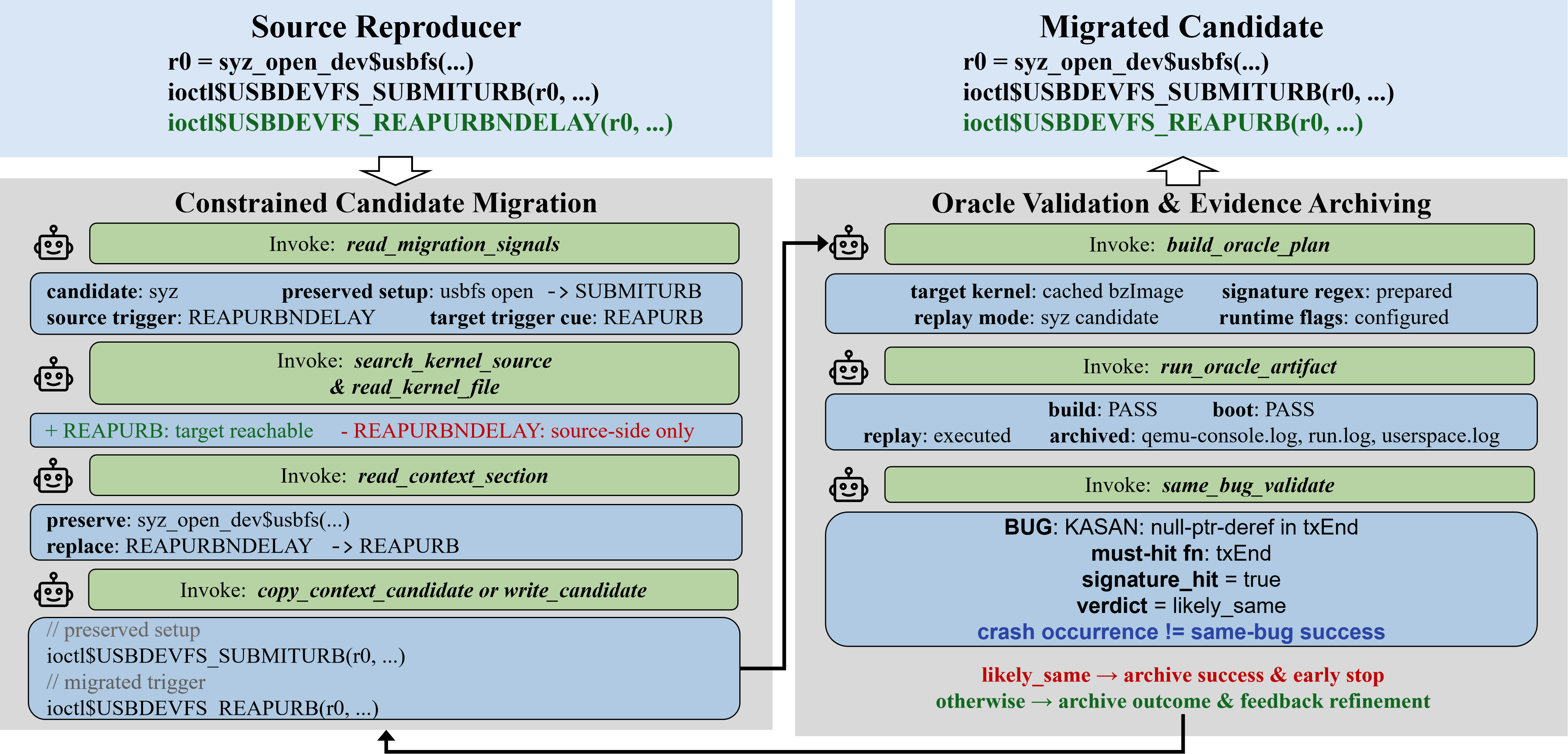}
\caption{Representative execution trace of constrained reproducer
migration and oracle-guided same-bug validation on the target kernel.}
\label{fig:migration-validation}
\end{figure*}

The migration stage transforms $X$ into an executable target-side
candidate. As illustrated in Figure~\ref{fig:migration-validation},
the agent operates in a bounded tool space over the structured context
and target kernel source, adapting version-sensitive behavior without
freely regenerating the entire crash reproducer.

\textbf{Evidence-guided adaptation.}
The LLM-based agent reads $S_{\mathrm{migration}}$ and $F_{\mathrm{repro}}$ to
identify operations that should remain stable and those likely
invalidated by kernel evolution. In the USBFS example of
Figure~\ref{fig:migration-validation}, it preserves the USBFS setup and
the \texttt{USBDEVFS\_SUBMITURB} sequence because they encode the
bug-triggering resource flow, then uses
\textsc{search\_kernel\_source} and \textsc{read\_kernel\_file} to
inspect the target implementation. The target source confirms that
\texttt{USBDEVFS\_REAPURB} remains reachable whereas
\texttt{USBDEVFS\_REAPURBNDELAY} is source-specific. The agent thus
preserves the setup skeleton and replaces only the version-sensitive
reap operation, preventing drift toward unrelated target-side
behavior.

\textbf{Bounded materialization.}
After selecting the adaptation, \tool{} materializes the candidate
through bounded file operations rather than unrestricted long-form
generation. The agent either copies an existing context candidate or
writes a localized edit with explicit provenance. In the running
example, the materialized syz candidate preserves the setup and URB
submission steps while replacing
\texttt{USBDEVFS\_REAPURBNDELAY} with
\texttt{USBDEVFS\_REAPURB}. This separation between semantic
adaptation and file materialization keeps edits small, records exactly
what changed, and avoids truncation or re-synthesis errors on large
reproducers. The full retry and failure-handling control flow for
candidate generation is included in the supplementary material.

\textbf{Constraint implementation.}
The generation constraints are implemented at two levels.
At the semantic level, prompt instructions require the agent to
preserve the bug-triggering setup, resource dependencies, and
reachable control flow while adapting only version-sensitive behavior.
At the operational level, candidates can be materialized only through
a bounded tool interface that enforces the action schema and
candidate-materialization contract. In particular, the interface
programmatically restricts the available actions, required arguments,
output locations, and candidate representation, whereas trigger
preservation remains a semantic instruction evaluated through
execution and oracle evidence.
Candidate executability and target replayability are subsequently
checked during validation, and observed failures are archived and
returned to the agent as feedback. These checks constrain how
candidates are constructed and validated, but do not provide a hard
guarantee that a generated candidate preserves the original bug
semantics.

\subsection{Execute and Validate}
\label{sec:oracle-validation}

The online oracle determines whether a target execution is sufficiently
consistent with the source bug to guide migration. It is used only for
feedback and early stopping, whereas final success is determined by
the offline confirmation protocol. As shown in
Algorithm~\ref{alg:validation-loop}, \tool{} iteratively executes and
refines candidates within the replay budget. An
\texttt{online\_same} outcome terminates the loop but does not constitute
the final success label reported in Section~\ref{sec:evaluation}.
The hidden target-reference reproducer, target-reference crash, and
fixing patch remain excluded from online validation and feedback.

\FloatBarrier
\begin{algorithm}[htbp]
\caption{Oracle-Guided Validation Loop in \tool{}}
\label{alg:validation-loop}
\footnotesize
\begin{algorithmic}[1]
\Require structured migration context $X$, target replay environment
$E$, replay budget $B$
\Ensure online outcome in
$\{\texttt{online\_same},\texttt{inconclusive}\}$

\For{$i \gets 1$ to $B$}
    \State $R \gets \Call{GetOrRefineCandidate}{X}$
    \If{$R = \varnothing$}
        \State \Return \texttt{inconclusive}
        \Comment{agent generation failed}
    \EndIf

    \State $T \gets \Call{TargetReplayExecution}{R,E}$

    \If{$T.\texttt{status}=\texttt{candidate\_invalid}$}
        \State \Call{ArchiveOutcome}
        {\texttt{candidate\_invalid},T.\texttt{evidence}}
        \State \Call{OracleGuidedFeedback}
        {X,T.\texttt{evidence}}
        \State \textbf{continue}
    \EndIf

    \If{$T.\texttt{status}\in
    \{\texttt{build\_unavailable},
      \texttt{replay\_interrupted},
      \texttt{timeout},
      \texttt{infrastructure\_uncertain}\}$}
        \State \Call{ArchiveOutcome}
        {T.\texttt{status},T.\texttt{evidence}}
        \State \Call{OracleGuidedFeedback}
        {X,T.\texttt{evidence}}
        \State \textbf{continue}
    \EndIf

    \If{$\neg T.\texttt{crash\_observed}$}
        \State \Call{ArchiveOutcome}
        {\texttt{oracle\_no\_crash},T.\texttt{evidence}}
        \State \Call{OracleGuidedFeedback}
        {X,T.\texttt{evidence}}
        \State \textbf{continue}
    \EndIf

    \State $V \gets
    \Call{OnlineSameBugValidation}
    {T.\texttt{evidence},X.\Omega_{\mathrm{online}}}$

    \If{$V.\texttt{status}=\texttt{same}$}
        \State \Call{ArchiveOutcome}
        {\texttt{online\_same},V.\texttt{evidence}}
        \State \Return \texttt{online\_same}
        \Comment{early stopping}
    \ElsIf{$V.\texttt{status}=\texttt{different}$}
        \State \Call{ArchiveOutcome}
        {\texttt{different\_bug},V.\texttt{evidence}}
        \State \Call{OracleGuidedFeedback}
        {X,V.\texttt{evidence}}
    \Else
        \State \Call{ArchiveOutcome}
        {\texttt{uncertain\_crash},V.\texttt{evidence}}
        \State \Call{OracleGuidedFeedback}
        {X,V.\texttt{evidence}}
    \EndIf
\EndFor

\State \Return \texttt{inconclusive}
\end{algorithmic}
\end{algorithm}

\textbf{Operational online oracle.}
The oracle normalizes the source crash signature and the observed
target crash into a tuple
$Z=\langle k,a,F,q\rangle$, where $k$ is the normalized failure family
and subtype, $a$ is the crash anchor, $F$ is the ordered sequence of
bug-relevant stack frames, and $q$ contains family-specific root-cause
cues. The normalizer strips instruction offsets and compiler-generated
suffixes such as \texttt{.isra}, \texttt{.constprop}, \texttt{.part},
and \texttt{.cold}. It filters generic reporting frames including
\texttt{kasan\_report}, \texttt{dump\_stack},
\texttt{warn\_slowpath}, and \texttt{panic}, and then compares only
the first eight remaining bug-relevant frames.

The online oracle uses a closed family system. Memory-safety crashes
must match the normalized subtype and have compatible access direction
and object-lifecycle cues when available. GPF, null-dereference, and
UBSAN crashes must preserve the fault class and faulting operation.
WARNING, BUG, and OOPS crashes must preserve the mapped assertion or
warning site. Hang and stall crashes must preserve the mapped terminal
or blocking point together with the blocked resource when available.
Deadlock and lockdep crashes must preserve the normalized lock pair or
dependency cycle and compatible acquisition context. The anchor must
map, and the required number of additional callers must also map in
order: at least one for memory-safety, GPF/null/UBSAN, and
WARNING/BUG/OOPS families, and at least two for hang/stall families.
Missing anchor evidence, ambiguous stack correspondence, or incomplete
family-specific evidence does not count as a match.

To compare source and target crashes across kernel revisions, the
oracle constructs an auditable function map $M$ using only three
mapping rules: exact normalized symbol equality, explicit function or
file correspondence derived from the source--target Git diff, and
inlining-aware mapping from a removed callee to its enclosing function
and call site. If multiple mappings remain possible, the oracle must
return \texttt{uncertain}. The procedure does not use fuzzy name
matching, embedding similarity, or scalar thresholds. Root-cause cues
in $q$ are treated symmetrically: if both crashes expose a cue and the
values conflict, the oracle returns \texttt{different}; if the cue is
missing or only partially observed, the oracle returns
\texttt{uncertain} unless the remaining evidence is sufficient under
the corresponding family rule.
Table~\ref{tab:oracle-rules} summarizes the mandatory evidence for an
online \texttt{same} decision. Each row defines both the minimum
positive evidence and the explicit conflict patterns that force a
\texttt{different} verdict. Executions that crash but fail to satisfy
either side conservatively remain \texttt{uncertain}.

\begin{table*}[t]
\centering
\caption{Failure-family rules for online same-bug validation.}
\label{tab:oracle-rules}
\scriptsize
\setlength{\tabcolsep}{5pt}
\renewcommand{\arraystretch}{1.15}
\renewcommand{\tabularxcolumn}[1]{m{#1}}

\rowcolors{2}{gray!10}{white}
\begin{tabularx}{\textwidth}{
@{}
>{\raggedright\arraybackslash}m{0.17\textwidth}
>{\raggedright\arraybackslash}m{0.43\textwidth}
>{\raggedright\arraybackslash}X
@{}
}
\toprule
\rowcolor{white}
\textbf{Failure family}
& \textbf{Requirements for \texttt{same}}
& \textbf{Conflicting evidence} \\
\midrule

Memory safety
& Same normalized subtype; mapped anchor and $\geq 1$ additional
caller; compatible access and object cues
& Subtype, access, object-lifecycle, or fault-site mismatch \\

GPF/null/UBSAN
& Same fault class; mapped faulting function or instruction and
$\geq 1$ additional caller
& Fault-class or faulting-operation mismatch \\

WARNING/BUG/OOPS
& Same mapped assertion or warning site and $\geq 1$ mapped caller
& Assertion-condition or warning-site mismatch \\

Hang/stall
& Same mapped blocking or terminal function and $\geq 2$ mapped
callers; compatible blocked resource when available
& Blocking-point, resource, or wait-condition mismatch \\

Deadlock/lockdep
& Same normalized lock pair or dependency cycle; compatible
acquisition sites
& Lock, dependency-cycle, or acquisition-context mismatch \\

\bottomrule
\end{tabularx}
\end{table*}

\begin{algorithm}[t]
\caption{Operational Online Same-Bug Oracle}
\label{alg:same-bug-oracle}
\footnotesize
\begin{algorithmic}[1]
\Require target evidence $e_t$, online oracle configuration
$\Omega_{\mathrm{online}}=\langle Z_s,M\rangle$
\Ensure verdict in
$\{\texttt{same},\texttt{different},\texttt{uncertain}\}$

\State $Z_t \gets \Call{NormalizeCrash}{e_t}$

\If{$Z_t=\varnothing$}
    \State \Return \texttt{uncertain}
\EndIf

\If{$\Call{ExplicitConflict}{Z_s,Z_t,M}$}
    \State \Return \texttt{different}
\EndIf

\If{$\neg\Call{MappedAnchorMatch}{Z_s.a,Z_t.a,M}$}
    \State \Return \texttt{uncertain}
\EndIf

\If{$\neg\Call{FamilyRuleSatisfied}{Z_s,Z_t,M}$}
    \State \Return \texttt{uncertain}
\EndIf

\State \Return \texttt{same}
\end{algorithmic}
\end{algorithm}

\textbf{Outcome archiving and oracle-guided feedback.}
An \texttt{online\_same} outcome stops refinement and sends the
archived execution to offline confirmation. Outcomes classified as
\texttt{different}, \texttt{uncertain}, \texttt{candidate\_invalid},
or \texttt{oracle\_no\_crash} are archived together with their replay
evidence and returned to the migration stage as the next-round
feedback signal. Build failures and infrastructure failures are
recorded separately and do not provide positive same-bug evidence. This
loop lets \tool{} refine candidates using execution-derived signals
without exposing the hidden target answer.

\section{Evaluation}
\label{sec:evaluation}
We evaluate \tool{} in two settings. The paired benchmark contains
107 same-bug revision pairs, each assigned a source--target orientation
for the main comparison. All methods receive source-side evidence and
the target kernel environment, while target-reference reproducers,
crashes, outcomes, and fixing patches are reserved for offline
evaluation.
The realistic-deployment setting
migrates historical crash reproducers from Linux~5.15, Linux~6.1, and
Linux~6.6 to Linux~7.2-rc3, where no target-reference reproducer or
reference crash is available. Candidate reproducers are executed on the
target kernel in QEMU, with build, boot, execution, and crash outcomes
recorded for validation and iterative refinement. Unless otherwise
stated, all main experiments use GPT-5.4 as the LLM backbone for both
\tool{} and the LLM-based baselines.

We evaluate \tool{} by answering the following four research questions:

\begin{itemize}[leftmargin=15pt]
    \item \textbf{RQ1: Overall effectiveness.} Can \tool{} successfully migrate historical Syzkaller crash reproducers across kernel versions and reproduce the same bug on the target kernel?

    \item \textbf{RQ2: Realistic deployment.} Can \tool{} generate target-side failure evidence suitable for upstream reporting when historical crash reproducers are migrated to an evolving kernel?

    \item \textbf{RQ3: Component contribution.} How much do the major design choices in \tool{} contribute to migration success?

    \item \textbf{RQ4: Model sensitivity and cost-effectiveness.} How sensitive is \tool{} to the underlying LLM backbone, and what tradeoffs arise across effectiveness, token usage and cost?
\end{itemize}

\subsection{Experimental Setup}
\label{Setup}

\begin{table*}[t]
\centering
\caption{Summary of the evaluation dataset.}
\label{tab:dataset-summary}
\scriptsize
\setlength{\tabcolsep}{4pt}
\renewcommand{\arraystretch}{1.08}

\begin{subtable}[t]{0.44\textwidth}
\centering
\caption{Artifact coverage and direct-replay baseline.}
\label{tab:dataset-overview}

\begin{tabularx}{\linewidth}{
>{\raggedright\arraybackslash}X
>{\centering\arraybackslash}p{0.12\linewidth}
>{\centering\arraybackslash}p{0.16\linewidth}
}
\toprule
\textbf{Artifact / Outcome}
& \textbf{Cases}
& \textbf{Coverage} \\
\midrule

syz reproducer
& 107 & 100.0\% \\

C reproducer
& 96 & 89.7\% \\

Both formats
& 96 & 89.7\% \\

Paired crash logs
& 107 & 100.0\% \\

Direct replay same-bug
& 19 & 17.8\% \\

\bottomrule
\end{tabularx}
\end{subtable}
\hfill
\begin{subtable}[t]{0.54\textwidth}
\centering
\caption{Distribution of migration pairs across kernel families.}
\label{tab:version-distribution}

\begin{tabular}{ccccc}
\toprule
& \textbf{Target 5.15}
& \textbf{Target 6.1}
& \textbf{Target 6.6}
& \textbf{Total} \\
\midrule

\textbf{Source 5.15}
& 5 & 0 & 48 & 53 \\

\textbf{Source 6.1}
& 0 & 23 & 26 & 49 \\

\textbf{Source 6.6}
& 1 & 1 & 3 & 5 \\

\midrule
\textbf{Total}
& 6 & 24 & 77 & \textbf{107} \\

\bottomrule
\end{tabular}
\end{subtable}

\end{table*}

\textbf{Dataset.}
\label{sec:dataset}
We construct the evaluation dataset from publicly available
Syzkaller/syzbot crash records. 
Rather than grouping crashes by title or faulting function alone, we
pair revisions through upstream--backport fixes implementing the same
logical change. This process yields 107 same-bug revision pairs across
Linux 5.15, 6.1, and 6.6. Each pair has a designated source--target
orientation for the main benchmark; RQ1 additionally evaluates the
eligible chronological directions according to target-side reproducer
availability.
Each case provides a source syz crash reproducer and a target execution
configuration, while 96 cases additionally provide a C crash reproducer.
Table~\ref{tab:dataset-summary}(a) summarizes artifact availability and
the direct-replay baseline outcome. Among the 107 cases, direct replay
confirms same-bug reproduction for only 19 cases (17.8\%), indicating
that historical crash reproducers often become ineffective when directly
transferred across kernel revisions. We use all 107 cases for the overall evaluation, component ablation, and model-sensitivity analysis.

\textbf{Baselines.}
We compare \tool{} with direct replay and two LLM-based reproducer
adaptation baselines. \emph{Direct replay} executes the original source
reproducer on the target kernel without modification.
\emph{LLM-DRA} prompts the same LLM used by \tool{} to generate a
target-side reproducer directly from the available source evidence and
target environment, without structured semantic enhancement,
domain-specific generation constraints, or target-side execution
feedback. \emph{LLM-DRA+Static} extends LLM-DRA with up to three rounds
of validation and self-repair using diagnostics from
\texttt{syz-prog2c -strict}. This feedback concerns only reproducer
validity and translation errors and does not reveal target-side crash
behavior. 
All methods are evaluated on the same cases and target environments
using the same replay procedure, execution timeout, and offline
confirmation protocol. 
Unless otherwise stated, \tool{} allows at most $B=3$
target-execution rounds per case, including the initial candidate.
The process terminates early once an \texttt{online\_same} outcome
is obtained.
The baselines represent direct reuse and
direct LLM-based adaptation with or without static validity feedback;
\tool{} additionally uses structured migration evidence and
target-execution feedback. Thus, this comparison evaluates complete
migration workflows rather than isolating the effect of any single
component. We examine component contributions separately in RQ3.

\textbf{Metrics.}
\label{sec:metrics}
Our primary metric is the number and percentage of cases achieving
confirmed same-bug migration. We define \emph{success} as a confirmed
same-bug migration and \emph{failure} as a case in which no such confirmation is obtained. We also report directional and family-level
results, token consumption, and cost per successful migration.
Final success is determined through blinded offline adjudication.
Executions classified as \texttt{online\_same} are independently
compared with the hidden target-reference crash by two kernel experts
using the failure-family-specific rules in
Table~\ref{tab:oracle-rules}. The review considers the normalized
failure subtype, mapped crash anchor, bug-relevant stack
correspondence, and available root-cause cues while withholding the
method name and online verdict. Each expert assigns \texttt{same},
\texttt{different}, or \texttt{uncertain}; disagreements are
adjudicated by a third expert. Only final \texttt{same} labels count as
successful migrations.
We separately evaluate the online oracle on a blinded, stratified set
of 60 crash-producing executions and report Cohen's $\kappa$,
precision, recall, F1, and decision coverage against the adjudicated
labels.

\subsection{RQ1: Overall Effectiveness}

We compare \tool{} with the three baselines described in
Section~\ref{Setup} on the same 107 source--target
migration cases. We measure overall effectiveness using the
\emph{confirmed same-bug migration rate}, defined as the
fraction of cases in which a method produces at least one candidate
receiving a final \texttt{same} label under the offline confirmation
protocol.
Direct replay executes the original source crash reproducer without
modification. LLM-DRA prompts the same LLM used by \tool{} to generate a
target-side reproducer directly from the available source evidence and
target environment, without semantic enhancement, domain-specific
constraints, or execution feedback. LLM-DRA+Static additionally allows
up to three rounds of validation and self-repair using diagnostics from
\texttt{syz-prog2c -strict}. This feedback concerns only syntactic and
translation validity and does not reveal target-side crash behavior.

As shown in Table~\ref{tab:overall-current}, \tool{} achieves 57
confirmed same-bug migrations (53.3\%), compared with 19 (17.8\%) for
direct replay, 18 (16.8\%) for LLM-DRA, and 17 (15.9\%) for
LLM-DRA+Static. Relative to direct replay, \tool{} improves the confirmed same-bug migration rate by 35.5 percentage points. It produces no non-executable candidate reproducers and leaves
no unresolved kernel build failures; its remaining 50 cases execute
without producing a target-side crash. By contrast, direct replay
produces 18 non-executable source crash reproducers and leaves 3 kernel build
failures, while LLM-DRA produces 43 non-executable reproducers and
leaves 5 kernel build failures. Static self-repair reduces the number
of non-executable reproducers from 43 to 28, but leaves 7 kernel build
failures, increases executions without a crash from 41 to 55, and does
not improve the confirmed same-bug migration rate. These results show that static validity repair alone does not improve
same-bug reproduction in the evaluated direct-generation workflow.
The complete \tool{} workflow obtains more confirmed migrations than
the evaluated baselines and resolves the target build failures
observed in this benchmark. The contribution of its individual
components is examined in RQ3.

Because \tool{} uses online oracle decisions for candidate refinement
and early stopping, we evaluate this signal on an independent set of
60 crash-producing executions. Two kernel experts independently label
each execution as \texttt{same}, \texttt{different}, or
\texttt{uncertain}, with disagreements adjudicated by a third expert.
The reviewers achieve a Cohen's $\kappa$ of 0.76 before adjudication.
Against the adjudicated labels, the online oracle achieves 92.0\%
precision, 81.0\% recall, an F1 score of 86.0\%, and 90.0\% decision
coverage for the \texttt{same} class. These measurements assess the
online feedback signal; migrations in
Table~\ref{tab:overall-current} are separately determined by the
offline confirmation protocol using hidden target-reference evidence.

\begin{table}[t]
\centering
\caption{Final migration outcomes on the 107-case benchmark.
A success denotes a confirmed same-bug migration.}
\label{tab:overall-current}
\scriptsize
\renewcommand{\arraystretch}{1.08}

\begin{tabular*}{\columnwidth}{
@{\extracolsep{\fill}}
lcccc
@{}
}
\toprule
\textbf{Method} &
\textbf{Success} &
\textbf{Executed Without Crash} &
\textbf{Reproducer Not Executable} &
\textbf{Kernel Build Failure} \\
\midrule

Direct replay
& 19/107 (17.8\%)
& 67
& 18
& 3 \\

LLM-DRA
& 18/107 (16.8\%)
& 41
& 43
& 5 \\

LLM-DRA+Static
& 17/107 (15.9\%)
& 55
& 28
& 7 \\

\tool{}
& 57/107 (53.3\%)
& 50
& 0
& 0 \\

\bottomrule
\end{tabular*}
\end{table}

\begin{table*}[t]
\centering

\begin{minipage}[t]{0.48\textwidth}
\centering
\caption{Forward-migration effectiveness of \tool{} on 104 eligible
source--target orientations.}
\label{tab:forward-migration}
\scriptsize
\setlength{\tabcolsep}{3.5pt}
\renewcommand{\arraystretch}{1.10}

\begin{tabular}{llccc}
\toprule
\textbf{Src. Family} &
\textbf{Tgt. Family} &
\textbf{Relation} &
\textbf{Succ./Eval.} &
\textbf{Rate} \\
\midrule

Linux 5.15
& Linux 5.15
& Within-family
& 5/5
& 100.0\% \\

Linux 6.1
& Linux 6.1
& Within-family
& 23/23
& 100.0\% \\

Linux 6.6
& Linux 6.6
& Within-family
& 3/3
& 100.0\% \\

Linux 5.15
& Linux 6.6
& Cross-family
& 19/47
& 40.4\% \\

Linux 6.1
& Linux 6.6
& Cross-family
& 7/26
& 26.9\% \\

\midrule
\multicolumn{3}{l}{\textbf{Overall}}
& \textbf{57/104}
& \textbf{54.8\%} \\

\bottomrule
\end{tabular}
\end{minipage}
\hfill
\begin{minipage}[t]{0.48\textwidth}
\centering
\caption{Backward-migration effectiveness of \tool{} on 79 eligible
source--target orientations.}
\label{tab:backward-migration}
\scriptsize
\setlength{\tabcolsep}{3.5pt}
\renewcommand{\arraystretch}{1.10}

\begin{tabular}{llccc}
\toprule
\textbf{Src. Family} &
\textbf{Tgt. Family} &
\textbf{Relation} &
\textbf{Succ./Eval.} &
\textbf{Rate} \\
\midrule

Linux 5.15
& Linux 5.15
& Within-family
& 5/5
& 100.0\% \\

Linux 6.1
& Linux 6.1
& Within-family
& 17/20
& 85.0\% \\

Linux 6.6
& Linux 6.6
& Within-family
& 3/3
& 100.0\% \\

Linux 6.6
& Linux 5.15
& Cross-family
& 15/32
& 46.9\% \\

Linux 6.6
& Linux 6.1
& Cross-family
& 8/19
& 42.1\% \\

\midrule
\multicolumn{3}{l}{\textbf{Overall}}
& \textbf{48/79}
& \textbf{60.8\%} \\

\bottomrule
\end{tabular}
\end{minipage}

\end{table*}

We further evaluate the 107 pairs by chronological direction. Let $A$
and $B$ denote the earlier and later revisions, respectively. A
direction is eligible when its target side has a reference reproducer
for final same-bug confirmation. Accordingly, 104 pairs support forward
migration ($A\!\rightarrow\!B$), while 79 support backward migration
($B\!\rightarrow\!A$). The two subsets overlap and are not independent:
the main 57/107 result follows the designated orientations in
Table~\ref{tab:dataset-summary}(b), whereas
Tables~\ref{tab:forward-migration} and
\ref{tab:backward-migration} use the corresponding eligible
chronological subsets. We classify a task as \emph{within-family} when
both revisions belong to the same major--minor family and as
\emph{cross-family} otherwise.

Tables~\ref{tab:forward-migration} and
\ref{tab:backward-migration} report the results for the two eligible
directional subsets. \tool{} achieves confirmed same-bug migration in
57 of the 104 eligible forward tasks (54.8\%) and 48 of the 79 eligible
backward tasks (60.8\%). These rates are not directly comparable:
the two subsets overlap but contain different cases because
target-side reproducer availability differs between the two
directions.
Within-family migration achieves 31/31
(100.0\%) in the forward direction and 25/28 (89.3\%) in the backward
direction, whereas the corresponding cross-family rates are 26/73
(35.6\%) and 23/51 (45.1\%), respectively. This consistent gap suggests that migration difficulty is closely related to the extent of source–target kernel evolution rather than chronological direction alone.

\textbf{Answer to RQ1:}
On the 107-case benchmark, \tool{} confirms 57 same-bug migrations
(53.3\%), compared with 19 for direct replay, 18 for LLM-DRA, and
17 for LLM-DRA+Static. The eligible forward and backward subsets
achieve 57/104 and 48/79 confirmed migrations, respectively; these
rates are based on different, overlapping subsets. The online oracle
achieves 92.0\% precision and 81.0\% recall.

\FloatBarrier

\subsection{RQ2: Deployment on an Evolving Kernel}

We start from 400 historical syzbot reports collected from the
Linux~5.15, 6.1, and 6.6 kernel families. We exclude seven reports
without a usable crash reproducer, leaving 393 migration inputs.
Target-side failures are observed for 82 inputs, while no target-side
failure is observed for the remaining 311 inputs. Among the 82
failure-producing inputs, 60 are triggered by the original crash
reproducers without adaptation and are therefore excluded from the
adaptation-focused deployment analysis. The remaining 22 failures
are produced by adapted candidates and are submitted as upstream
reports after deduplication and manual triage.

Unlike RQ1, this deployment does not support reference-based same-bug
confirmation. We therefore evaluate whether the adapted candidates
produce target-side failure evidence suitable for upstream reporting,
and use maintainer confirmation and upstream fixes as external evidence
of practical relevance. We classify the community outcomes as follows.
\emph{Reported} denotes a submitted report that has not yet received
a definitive resolution; \emph{Confirmed} denotes explicit maintainer
acknowledgement of the reported issue; and \emph{Fixed} denotes that
a corresponding fix has been accepted upstream and verified to address
the reported issue.

\begin{table*}[htbp]
\centering
\caption{Community outcomes of 22 crash reports generated through
cross-version crash reproducer migration on Linux~7.2-rc3, including
7 fixed and 5 maintainer-confirmed issues.}
\label{tab:rc3-candidates}
\scriptsize
\setlength{\tabcolsep}{6pt}
\renewcommand{\arraystretch}{0.98}

\begin{tabular}{@{}c l c l c@{}}
\toprule
\textbf{Idx} &
\textbf{Crash ID} &
\textbf{Source family} &
\textbf{Target-side failure} &
\textbf{Status} \\
\midrule

1
& \texttt{0fb22d8d1996d2661445}
& 5.15
& WARNING in \texttt{gl861\_ctrl\_msg}
& Reported \\

2
& \texttt{60b5e90b6d2c52328fb9}
& 5.15
& Task hung in \texttt{gfs2\_jhead\_process\_page}
& Reported \\

3
& \texttt{751290054fe23f2a5317}
& 5.15
& Task hung in \texttt{\_\_gfs2\_lookup}
& Reported \\

4
& \texttt{80628191ce3caee0ba98}
& 5.15
& General protection fault in \texttt{device\_move}
& Reported \\

5
& \texttt{80ea31bb816859827846}
& 5.15
& WARNING in \texttt{dtv5100\_i2c\_msg}
& Reported \\

6
& \texttt{8c223a3aa6865f107658}
& 5.15
& Possible circular locking dependency in \texttt{ext4\_writepages}
& Reported \\

7
& \texttt{1782a3a587bc53dc10cf}
& 5.15
& possible deadlock in \texttt{ocfs2\_write\_begin\_nolock}
& Fixed \\

8
& \texttt{1050c0099ec5bfe7ee4e}
& 5.15
& possible deadlock in \texttt{mdc800\_device\_open}
& Fixed \\

9
& \texttt{83e39b821587064bd7a3}
& 5.15
& general protection fault in \texttt{set\_link\_state}
& Confirmed \\

10
& \texttt{91209a63bd6e8b2608e4}
& 6.1
& KASAN: use-after-free read in \texttt{hpfs\_get\_ea}
& Confirmed \\

11
& \texttt{1c0f13d7bc65d06d5fc9}
& 6.6
& RCU stall in \texttt{addrconf\_rs\_timer}
& Fixed \\

12
& \texttt{3aec4eac211b2b7142b7}
& 6.6
& Kernel page fault in \texttt{udf\_add\_fid\_counter}
& Reported \\

13
& \texttt{35ad01f7f3622937dc14}
& 6.6
& WARNING in \texttt{\_\_rate\_control\_send\_low}
& Reported \\

14
& \texttt{35c6202df82fb8c32975}
& 6.6
& Possible circular locking dependency in \texttt{usbg\_make\_tpg}
& Confirmed \\

15
& \texttt{420b1c24d8512ae0410c}
& 6.6
& WARNING in \texttt{free\_large\_kmalloc}
& Confirmed \\

16
& \texttt{5b6de2def31c29d5a343}
& 6.6
& Non-static lock-key warning in \texttt{ocfs2\_dlm\_shutdown}
& Reported \\

17
& \texttt{7ad36bea639f2d542633}
& 6.6
& NULL-pointer dereference in \texttt{nvmet\_file\_ns\_enable}
& Reported \\

18
& \texttt{7bec276b8b8889ad21ab}
& 6.6
& WARNING in \texttt{ceph\_con\_v2\_try\_read}
& Fixed \\

19
& \texttt{8efdc8e70c042d84d068}
& 6.6
& WARNING in \texttt{fanotify\_handle\_event}
& Fixed \\

20
& \texttt{d851455dbf571fe8e225}
& 6.6
& WARNING in \texttt{usb\_ep\_enable}
& Fixed \\

21
& \texttt{e6aa07b89235491b9670}
& 6.6
& General protection fault in \texttt{dummy\_set\_selfpowered}
& Confirmed \\

22
& \texttt{5916dc0ad70b3adbefcd}
& 6.6
& KASAN use-after-free in \texttt{rc\_dev\_uevent}
& Fixed \\

\bottomrule
\end{tabular}
\end{table*}

The 22 submitted reports were generated from nine source cases on Linux 5.15, one source case on Linux 6.1, and twelve source cases on Linux 6.6.
Among the 22 reports, 7 issues have been fixed, 5 have received explicit maintainer confirmation, and the remaining 10 have not yet received a definitive resolution. Each row in Table~\ref{tab:rc3-candidates}
corresponds to one submitted report generated from a distinct source case and records its community outcome.
We do not infer distinct root causes beyond the evidence provided by
maintainer responses or corresponding fixes.
These results show that \tool{} can transfer historical
crash-triggering knowledge to an evolving upstream kernel without
access to an existing target-side reproducer. More importantly, the
resulting executions provide sufficient evidence for practical kernel
bug reporting: twelve of the 22 reports had already received explicit
community validation through maintainer confirmation or upstream fixes.

\textbf{Illustrative Case: Resource-Semantic Adaptation and
Validation Boundary.}
A paired-benchmark case illustrates why target-side reachability alone
does not establish same-bug migration. Kernel evolution changes both
the BPF map type and the relevant tracepoint.
Listing~\ref{lst:resource-adaptation} shows how \tool{} adapts
\texttt{SOCKHASH} to \texttt{SOCKMAP} and redirects the attachment from
\texttt{mm\_page\_alloc} to \texttt{contention\_end}.

\Needspace{12\baselineskip}
\begin{lstlisting}[
    style=syzstyle,
    breaklines=true,
    escapeinside={(*@}{@*)},
    caption={Resource and tracepoint adaptation performed by \tool{}.},
    label={lst:resource-adaptation},
    captionpos=t,
    aboveskip=6pt,
    belowskip=6pt
]
// [Source] mm_page_alloc + SOCKHASH
bpf$MAP_CREATE(0x0, &@base={0x12, 0x4, 0x8, 0x8}, 0x48)
bpf$BPF_RAW_TRACEPOINT_OPEN(..., "(*@{\color{deepblue}\bfseries\ttfamily mm\_page\_alloc}@*)", r1, ...)

// [Adapted Candidate] contention_end + SOCKMAP
bpf$MAP_CREATE(0x0, &@base={0xf, 0x4, 0x8, 0x8}, 0x48)
bpf$BPF_RAW_TRACEPOINT_OPEN(..., "(*@{\color{deepred}\bfseries\ttfamily contention\_end}@*)", r1, ...)
\end{lstlisting}

The adaptation changes two version-sensitive elements while retaining
the structure of the original trigger. First, the BPF map type changes
from \texttt{SOCKHASH} (\texttt{0x12}) to \texttt{SOCKMAP}
(\texttt{0xf}), preserving the socket-map resource semantics required
by the BPF program. Second, the attachment point changes from
\texttt{mm\_page\_alloc} to \texttt{contention\_end}, providing a
target-compatible tracepoint through which the adapted program can
reach \path{sock_map_delete_elem}. The remaining BPF program structure
is preserved, and the adapted candidate has the same BPF program tag
as the hidden target-reference reproducer. These observations indicate
that the candidate reaches the intended target-side resource operation,
but they do not yet establish that it reproduces the same bug.

\begin{lstlisting}[
    style=syzstyle,
    breaklines=true,
    aboveskip=4pt,
    belowskip=4pt,
    caption={Failure manifestations of the source crash reproducer, target-reference reproducer, and adapted candidate.},
    label={lst:failure-manifestations}
]
// [Source, Linux 5.15] mm_page_alloc + SOCKHASH
__local_bh_enable_ip                 <- WARN (softirq.c:363)
  sock_hash_delete_elem
    bpf_trace_run4
      trace_mm_page_alloc
// [Target Reference, Linux 6.1] contention_end + SOCKMAP
HARDIRQ-safe -> HARDIRQ-unsafe lock order
  sock_map_delete_elem
    bpf_trace_run2
      __traceiter_contention_end
        <- in rcu_report_exp_cpu_mult
// [Adapted Candidate, Linux 6.1] contention_end + SOCKMAP
__local_bh_enable_ip                 <- WARN (softirq.c:376)
  sock_map_delete_elem
    bpf_trace_run2
      __traceiter_contention_end
        <- in do_exit
\end{lstlisting}

Listing~\ref{lst:failure-manifestations} shows why this path-level
correspondence is insufficient. The adapted candidate and the target
reference both execute \path{sock_map_delete_elem} through
\path{contention_end}, but their failures differ in both subtype and
execution context. The target reference reports a
\emph{HARDIRQ-safe--HARDIRQ-unsafe lock-order} violation reached from
\path{rcu_report_exp_cpu_mult}, whereas the adapted candidate produces
a warning in \path{__local_bh_enable_ip} during
\path{do_exit}. The adapted candidate is therefore path-aligned with the target reference but does not provide sufficient evidence of the same underlying failure.
This case demonstrates the validation boundary of reproducer migration:
preserving the program body and reaching the intended target-side
operation are necessary but not sufficient for confirming migration
success. A target-side crash must additionally exhibit failure evidence
consistent with the original bug under the case-specific same-bug
oracle. Accordingly, because the Linux~7.2-rc3 deployment has no hidden
target-reference crashes, we report its outputs as upstream crash
submissions rather than confirmed same-bug migrations, and use
maintainer confirmation or upstream fixes as external evidence of
practical relevance.

\textbf{Answer to RQ2:}
Of 393 migration inputs, 82 produce target-side failures: 60 without
adaptation and 22 through adapted candidates. All 22 adapted failures
are submitted upstream after deduplication and manual triage; seven
have been fixed, five have been maintainer-confirmed, and ten remain
unresolved. These outcomes provide external evidence of practical
relevance without establishing same-bug migration for unresolved cases.

\FloatBarrier
\subsection{RQ3: Component Contribution}

\tool{} consists of three major components supporting different stages
of cross-version migration. First, \emph{multi-source semantic
enhancement} integrates the source crash report, original reproducer,
bug-relevant resource dependencies, source and target kernel code, and
cross-version interface and execution-path differences into a
structured migration context. Second, the \emph{domain-specific
constrained agent} generates migration candidates under kernel-specific
transformation rules and bounded tool operations, thereby limiting
unnecessary modifications to the original trigger. Third,
\emph{oracle-guided iterative refinement} executes each candidate,
classifies its outcome, and returns case-specific diagnostic evidence
to guide subsequent adaptation.

We evaluate three ablated variants.
\tool{}-$S$ removes structured semantic enhancement while retaining
constrained generation and execution feedback. \tool{}-$C$ removes
the domain-specific semantic constraints from the generation prompt
while retaining the tool interface required for candidate
materialization. \tool{}-$O$ disables oracle-guided feedback and
iterative refinement while retaining semantic enhancement and
constrained generation.

\begin{table}[t]
\centering
\caption{Component ablation results on the 107-case benchmark.
$\Delta$ denotes the difference in the number of confirmed same-bug migrations relative to the full \tool{} pipeline.}
\label{tab:ablation-transition}
\scriptsize
\setlength{\tabcolsep}{3pt}
\renewcommand{\arraystretch}{1.15}

\begin{tabular}{lccc|cc}
\toprule
\textbf{Variant} &
\shortstack{\textbf{Semantic Enhancement}} &
\shortstack{\textbf{Generation Constraints}} &
\shortstack{\textbf{Oracle-Guided Feedback}} &
\textbf{Success} &
\shortstack{\textbf{$\Delta$ vs Full}} \\
\midrule

\tool{}
& \checkmark
& \checkmark
& \checkmark
& 57/107 (53.3\%)
& 0 \\

\tool{}-$S$
& \shortstack{No semantic extraction}
& \checkmark
& \checkmark
& 39/107 (36.4\%)
& -18 \\

\tool{}-$C$
& \checkmark
& \shortstack{Unconstrained generation}
& \checkmark
& 36/107 (33.6\%)
& -21 \\

\tool{}-$O$
& \checkmark
& \checkmark
& \shortstack{No feedback (one-shot)}
& 42/107 (39.3\%)
& -15 \\

\bottomrule
\end{tabular}
\end{table}
All variants use the same 107 cases, target environments, replay
procedure, and offline confirmation protocol. As shown in
Table~\ref{tab:ablation-transition}, removing semantic enhancement, generation
constraints, or oracle-guided refinement reduces confirmed migrations
from 57 to 39, 36, and 42, respectively. The largest degradation occurs
without generation constraints, suggesting that unrestricted adaptation
is more likely to alter behavior essential to the original trigger.
The \tool{}-$O$ comparison measures the deployed refinement loop as a
whole, including its additional candidate attempts, rather than the
isolated effect of oracle feedback under an equal candidate budget.

\textbf{Answer to RQ3:}
Removing semantic enhancement, generation constraints, or
feedback-guided refinement reduces the confirmed migration rate from
53.3\% to 36.4\%, 33.6\%, and 39.3\%, respectively, supporting all
three components within the complete \tool{} workflow.

\FloatBarrier

\subsection{RQ4: Model Sensitivity and Cost-Effectiveness}

We evaluate the sensitivity of \tool{} to the underlying LLM backbone
by varying only the model while keeping the migration workflow, tool
interface, oracle policy, replay budget, and 107-case benchmark fixed.
We compare migration success and failure, average input and output
tokens per case, and model cost per successful migration. A failure
denotes a case in which the model does not produce a confirmed same-bug
migration. Model costs are calculated from recorded token usage using
the corresponding API prices at the time of the experiments and exclude
kernel-execution and infrastructure costs. Holding the surrounding
pipeline and evaluation cases fixed allows us to isolate differences
associated with the LLM backbone.

\begin{table*}[t]
\centering
\caption{Model sensitivity and cost-effectiveness on the 107-case
benchmark. A failure denotes a case without a confirmed same-bug
migration. All models use the same \tool{} pipeline, oracle policy, and
replay budget. Token counts are averages per case.}
\label{tab:model-template}
\scriptsize
\setlength{\tabcolsep}{4pt}
\renewcommand{\arraystretch}{1.08}

\begin{tabular*}{\textwidth}{
@{\extracolsep{\fill}}
lccccc
}
\toprule
\textbf{Model} &
\textbf{Success} &
\textbf{Failure} &
\textbf{Avg. input token} &
\textbf{Avg. output token} &
\textbf{Cost / succ. (\$)} \\
\midrule

GPT-4o-mini
& 36/107 (33.6\%)
& 71
& 46{,}722
& 2{,}348
& 0.0862 \\

GPT-5
& 48/107 (44.9\%)
& 59
& 58{,}558
& 1{,}382
& 0.2507 \\

GPT-5.4
& 57/107 (53.3\%)
& 50
& 54{,}836
& 1{,}523
& 0.3207 \\

DeepSeek-V4-Pro
& 44/107 (41.1\%)
& 63
& 149{,}087
& 7{,}150
& 0.3194 \\

\bottomrule
\end{tabular*}
\end{table*}

Table~\ref{tab:model-template} shows substantial variation across LLM
backbones under the same migration workflow. GPT-5.4 achieves the highest confirmed migration rate, successfully migrating 57 of the 107 cases
(53.3\%). GPT-5 successfully migrates 48 cases (44.9\%), followed by
DeepSeek-V4-Pro with 44 cases (41.1\%) and GPT-4o-mini with 36 cases
(33.6\%). Thus, GPT-5.4 improves the success rate by 8.4 percentage
points over GPT-5, 12.2 percentage points over DeepSeek-V4-Pro, and
19.7 percentage points over GPT-4o-mini. These results indicate that
backbone choice substantially affects the agent's ability to interpret
structured migration evidence, interact with constrained tools, and
produce candidates that reproduce the original bug.

The cost results reveal an effectiveness--efficiency trade-off.
GPT-4o-mini has the lowest monetary cost per successful migration
(\$0.0862), but also achieves the lowest success rate. GPT-5 provides
an intermediate trade-off, achieving 48 successful migrations at
\$0.2507 per success. GPT-5.4 achieves the highest success rate at
\$0.3207 per successful migration. DeepSeek-V4-Pro incurs a comparable
cost of \$0.3194 per success while producing 13 fewer successful
migrations than GPT-5.4. Thus, GPT-5.4 provides the highest
effectiveness, GPT-4o-mini minimizes monetary cost, and GPT-5 offers a
middle ground between them.
Token consumption further differentiates the backbones. GPT-4o-mini,
GPT-5, and GPT-5.4 consume approximately 47K--59K input tokens per
case, whereas DeepSeek-V4-Pro consumes approximately 149K, more than
2.5$\times$ as many as any other model. Its average output length
reaches 7{,}150 tokens per case, compared with approximately
1{,}400--2{,}300 for the other backbones. Despite this substantially
larger token footprint, DeepSeek-V4-Pro does not achieve higher
migration effectiveness, indicating that greater token consumption
alone does not produce better migration outcomes.

\textbf{Answer to RQ4:}
GPT-5.4 achieves the highest migration rate, GPT-4o-mini minimizes
monetary cost, and GPT-5 provides an intermediate trade-off.
DeepSeek-V4-Pro consumes substantially more tokens without matching
GPT-5.4's effectiveness.

\FloatBarrier

\section{Threats to Validity}
\label{sec:threats}

\textbf{Construct validity.}
Confirmed same-bug migration is an operational approximation of bug
identity rather than direct proof of an identical root cause. We
mitigate this threat by combining normalized failure characteristics,
cross-version crash-anchor correspondence, bug-relevant stack context,
and family-specific root-cause cues, rather than relying on crash
occurrence or function-name similarity alone. Final effectiveness
results are determined through blinded offline adjudication using
hidden target-reference evidence, while the online oracle is evaluated
separately on an independently labeled set. Nevertheless, semantic
differences that are not observable from the available crash evidence
may still lead to misclassification.

\textbf{Internal validity.}
Execution and LLM nondeterminism may introduce variation across runs.
We mitigate this threat by applying consistent replay settings,
execution budgets, and confirmation criteria across all evaluated
methods and variants. The main baseline comparison evaluates complete
workflows that differ in their use of structured context and
target-execution feedback. Likewise, the one-shot \tool{}-$O$ ablation
changes both feedback availability and candidate opportunities. We
therefore interpret these comparisons as workflow-level evidence rather
than estimates of the isolated effect of a single mechanism.

\textbf{External validity.}
Our 107-case benchmark covers three Linux kernel families and requires
target-reference evidence for offline confirmation. It therefore
contains known-positive target cases and evaluates migration under
confirmable conditions rather than the prevalence of affected revisions
in the wild. The results may not generalize to other architectures,
vendor kernels, or substantially different crash artifacts. We mitigate
this limitation through within-family, cross-family, forward, backward,
and Linux~7.2-rc3 deployment evaluations; deployment outcomes may also
evolve as maintainers continue their investigations.

\section{Related Work}
\label{sec:related}

\textbf{Kernel Fuzzing and LLM-Assisted Input Generation.}
Kernel fuzzing research primarily focuses on discovering previously
unknown defects. Syzkaller and syzbot provide widely used
coverage-guided fuzzing and crash-collection infrastructure for the
Linux kernel~\cite{Syzkaller,Syzlang,syzbot}. Subsequent systems improve
seed selection and syscall modeling through trace distillation,
relation learning, specification generation, and reinforcement
learning~\cite{pailoor2018moonshine,sun2021healer,sun2022ksg,SyzVegas}.
Other work explores hardware-assisted kernel fuzzing, driver-interface
modeling, concurrency and file-system testing, and directed exploration
of selected program locations~\cite{schumilo2017kafl,corina2017difuze,
jeong2019razzer,xu2019janus,xu2020krace,Hawkeye,tan2023syzdirect}.
More recently, Fuzz4All and KernelGPT use LLMs to generate test inputs
or kernel specifications~\cite{xia2024fuzz4all,yang2025kernelgpt}.
These techniques seek to discover new behaviors or improve fuzzing
coverage. In contrast, \tool{} starts from an existing crash
reproducer and adapts it across kernel versions while preserving the
identity of the original defect.

\textbf{Failure Reproduction and Test-Case Reduction.}
Failure-reproduction techniques reconstruct executable failures from
recorded program states, field execution information, or natural
language bug reports. ReCrash preserves object states for replay,
BugRedux synthesizes executions that reproduce field failures, and
LIBRO uses LLMs to generate tests from bug reports
~\cite{artzi2008recrash,jin2012bugredux,kang2023libro}. In the kernel
domain, YOME applies reverse fuzzing to reproduce known Linux kernel
bugs, while SyzScope analyzes the security impact of fuzzer-exposed
crashes~\cite{li2025yome,zou2022syzscope}. Complementary work on delta
debugging, C-Reduce, and Perses minimizes failure-inducing inputs while
retaining an observed failure
~\cite{zeller2002deltadebugging,regehr2012creduce,sun2018perses}.
These approaches reconstruct, analyze, or minimize failures in a given
execution environment. \tool{} instead addresses the orthogonal
problem of transferring an already available crash reproducer to a different
kernel revision, where interface, configuration, and resource-semantic
changes may invalidate the original trigger.

\textbf{Crash Equivalence and Root-Cause Matching.}
Determining whether two executions represent the same defect is closely
related to crash deduplication. ReBucket clusters crash reports using
call-stack similarity, Igor performs root-cause-oriented crash
clustering, and FuzzerAid derives fault signatures for grouping fuzzing
crashes~\cite{dang2012rebucket,jiang2021igor,joshy2022fuzzeraid}.
These techniques primarily organize already observed failures and
reduce duplicate reports. \tool{} uses related evidence for a different
purpose: crash anchors, normalized stack correspondence, failure
subtypes, and root-cause cues are incorporated into a case-specific
oracle that guides candidate refinement and checks whether a migrated
execution retains the intended bug identity across kernel versions.

\textbf{Test Evolution, Repair, and Agentic Adaptation.}
Prior work repairs tests invalidated by software evolution, including
changes to method declarations, interfaces, and expected behavior
~\cite{mirzaaghaei2010repairing,mirzaaghaei2014test,
daniel2009reassert}. Automated patch-porting techniques similarly adapt
code changes across diverged repositories~\cite{pan2024patchporting}.
More broadly, SWE-agent, AutoCodeRover, and RepairAgent combine LLMs
with repository inspection, external tools, and execution feedback for
software maintenance and program repair
~\cite{yang2024sweagent,zhang2024autocoderover,
bouzenia2025repairagent}. Reproducer migration differs from conventional
test or program repair because producing a syntactically valid program or completing a target-side execution is insufficient: the adapted program
must continue to trigger the original failure. \tool{} specializes
agentic adaptation for this requirement by constructing structured
cross-version evidence, constraining version-sensitive edits, and using
target execution feedback for online refinement while reserving final
same-bug confirmation for independent offline evaluation.

\section{Conclusion}
\label{sec:conclusion}

This paper presents \tool{}, a target-aware framework for migrating
Linux kernel crash reproducers across revisions while preserving
same-bug evidence. By distinguishing bug-preserving behavior from
version-sensitive assumptions, \tool{} combines structured
cross-version context, constrained adaptation, and execution-guided
validation to transform crash reproducer reuse from direct replay into
an iterative migration process. Across 107 real-world cases, \tool{}
confirms 57 same-bug migrations, compared with 19 using direct replay,
and supports both forward and backward migration across kernel
revisions and families. Its deployment on Linux~7.2-rc3 further
produces 22 upstream reports, including 7 fixed and 5
maintainer-confirmed issues. These results show that successful
cross-version migration requires more than restoring executability:
adaptation must preserve the original trigger semantics, and the
resulting failure must be validated against bug-identity evidence.
Overall, \tool{} provides a practical foundation for reusing historical
crash reproducers in regression testing, bug revalidation, and
cross-version impact analysis.


\bibliographystyle{unsrt}
\bibliography{software}

\appendix

\end{document}